\documentclass[english,format=sigconf, review=false, screen=true, nonacm]{acmart}
\PassOptionsToPackage{ruled}{algorithm2e}
\usepackage[T1]{fontenc}
\usepackage[latin9]{inputenc}
\usepackage{cprotect}
\usepackage{refstyle}
\usepackage{algorithm2e}
\usepackage{graphicx}
\usepackage{tablefootnote}

\makeatletter

\newcommand{\noun}[1]{\textsc{#1}}
\providecommand{\tabularnewline}{\\}
\RS@ifundefined{subsecref}{
  \newref{subsec}{
        name      = \RSsectxt,
        names     = \RSsecstxt,
        Name      = \RSSectxt,
        Names     = \RSSecstxt,
        refcmd    = {\S\ref{#1}},
        rngtxt    = \RSrngtxt,
        lsttwotxt = \RSlsttwotxt,
        lsttxt    = \RSlsttxt
  }
}{}

\SetAlFnt{\small}
\SetAlCapFnt{\small}
\SetAlCapNameFnt{\small}
\SetAlCapHSkip{0pt}
\IncMargin{-\parindent}

\usepackage{listings}
\lstdefinelanguage{Rust}{
    sensitive=true,
    morekeywords=[1]{
        as, break, const, continue, crate, else, enum, extern,
        false, fn, for, if, impl, in, let, loop, match, mod,
        move, mut, pub, ref, return, self, Self, static, struct,
        super, trait, true, type, unsafe, use, where, while,
        async, await, dyn, abstract, become, do, final, macro,
        override, priv, typeof, unsized, virtual, yield
    },
    morekeywords=[2]{
        i8, i16, i32, i64, i128, isize,
        u8, u16, u32, u64, u128, usize,
        f32, f64, bool, char, str, String, Option, Result
    },
    morecomment=[l]{//},
    morecomment=[s]{/*}{*/},
    morestring=[b]",
    morestring=[b]'
}

\lstdefinelanguage{Prela}{
    sensitive=true,
    string=[d]{"},
    morekeywords={
       with, and, s, eq, select, gt, lt, filt, reduce, fold, inv, group_by, rx, map, gather, drive, probe, for, in, len, drv, prb
    }
}

\lstdefinelanguage{rql}[]{SQL}{
  deletekeywords={year}
}

\renewcommand{\noun}[1]{\textsf{#1}}

\usepackage{newtxmath}

\usepackage{forest}
\usepackage{tikz}
\usetikzlibrary{arrows.meta}

\ifdefined\showcaptionsetup
 \PassOptionsToPackage{caption=false}{subfig}
\fi
\usepackage{subfig}
\makeatother

\usepackage{babel}
\usepackage{listings}
\usepackage{listings}
\begin{document}
\title{Revisiting the Algebraic Foundation of Relational Data}
\author{Yisu Remy Wang}
\affiliation{\institution{University of California, Los Angeles} \department{Department of Computer Science}
\city{Los Angeles, CA}\country{USA}}
\author{Paul Talma}
\affiliation{\institution{University of California, Los Angeles} \department{Department of Philosophy}
\city{Los Angeles, CA}\country{USA}}
\begin{abstract}
We revisit Tarski's Algebra of Relations (\noun{TAR}), an old formalism
of relations predating Codd's relational algebra by over 100 years,
as a new foundation for relational databases. We argue \noun{TAR}
provides a better abstraction at both the semantic level and the physical
level, in the context of modern application code and system architecture.
To demonstrate the strengths of \noun{TAR}, we design and implement
Prela, a compositional and controllable query language, and show that
queries written in Prela are concise, clear, and efficient. 
\end{abstract}
\maketitle

\section{Introduction\label{sec:Introduction}}

Over the past 50 years, Codd's relational model\,\citep{DBLP:journals/cacm/Codd70}
has cemented its status as the foundation underlying database systems.
The relational algebra has therefore become the central abstraction
bridging high-level semantics to low-level execution. However, the
world has changed both below and above this abstraction. At the system
level, OLAP engines increasingly adopt column-oriented storage, while
transactional systems are built upon a key-value core. On the application
side, researchers are pushing towards higher abstractions like the
Entity/Relationship model\,\citep{DBLP:conf/cidr/Deshpande25,DBLP:journals/tods/Chen76},
with the goal to better support nested and graph data, and to reconcile
the long-standing impedance mismatch between relational schema and
application code. As a result, both the physical layer and the semantic
layer have drifted away from the original table-oriented abstraction.
Yet, as if by serendipity, the independent evolution of the physics
and the semantics of relational data have led to the same destiny!
This reunion calls for new abstractions that better connect queries
to their execution. In this paper, we revisit an old formalism of
relations predating Codd's relational algebra by over 100 years. This
formalism was developed by logicians from De Morgan, Peirce, Schröder,
to Tarski and Givant, and is now commonly know as Tarski's Algebra
of Relations (\noun{TAR})\,\citep{DBLP:journals/jsyml/Tarski41,tarski1987formalization,Maddux1991-MADTOO}.
To demonstrate the potential of \noun{TAR} as a new foundation for
modern database systems, we design and implement Prela, a \emph{compositional}
and \emph{controllable} query language. Using Prela, the programmer
can write concise, clear, and efficient queries, all the while retaining
fine-grained control over low-level details of query execution. 

\begin{figure}
\begin{lstlisting}[language=Prela]
movie.with(company.s(country).eq("[us]")
      .and(keyword.eq("character-name-in-title"))
   .select(title
      .and(cast.s(person).s(alias).s(text)))
\end{lstlisting}

\caption{\label{fig:JOB-query-16b}JOB query 16b in Prela}
\end{figure}

The power of Prela is best illustrated with an example. Consider query
16b from the Join Order Benchmark\,\citep{DBLP:journals/pvldb/LeisGMBK015},
whose SQL formulation spans 23 clauses, 11 of which are used to specify
join conditions. \Figref{JOB-query-16b} shows the complete Prela
source for the same query. We will revisit this query in detail later,
but the intent is already clear at a glance: it looks for movies made
by American companies with keyword matching ``character-name-in-title'',
then outputs the movie title along with the alias for each cast member.
There are several notable differences between the Prela query and
the SQL version. First, there is no explicit \lstinline[language=SQL]!FROM!
clause in Prela; instead, relations appear as arguments to \emph{relation
combinators} like \lstinline!.with!, \lstinline!.and!, and \lstinline!.select!,
revealing the \emph{structure} of the query. Similarly, join conditions
are not specified by explicit join clauses, but by query structure.
Finally, thanks to Prela's algebraic nature, queries can be written
in a compositional way: in the example, every subexpression, including
\lstinline!company.s(country).eq("[us]")!, \lstinline!keyword.eq(...)!,
and \lstinline!cast.s(person).s(alias).s(text)!, is on its own a
valid query, and they appear directly as inputs to relation combinators
to build up the larger query. 

The reader may be surprised to learn that the query in \figref{JOB-query-16b}
is in fact regular Rust code. Indeed, \lstinline!movie!, \lstinline!company!,
\lstinline!country!, etc., are Rust variables, and combinators like
\lstinline!.with!, \lstinline!.and!, \lstinline!.select!, etc.,
are Rust functions. In other words, Prela is an \emph{embedded query
language}, or more simply, a \emph{library}. This means Prela queries
seamlessly integrate with application code, and the programmer can
pass in regular Rust functions as UDFs. Under the hood, every Prela
query is also a query plan, since each relational combinator corresponds
precisely to an operator in \noun{TAR}. A common complaint in SQL
is the difficulty of guiding the query optimizer when it fails to
find an efficient plan; by contrast, the Prela user has complete control
over all aspect of query planning, including join ordering, operator
pushdown, materialization, and selection of physical data structures.
This level of control is increasingly relevant in the age of AI, as
it puts the user---human or machine---back in the driver seat of
performance tuning. Nevertheless, this great power need not come with
great responsibility, and Prela/\noun{TAR} can just as well serve
as the \emph{intermediate representation} underlying a query optimizer---the
implementation of which we leave as future work. Finally, Prela queries
are surprisingly ``closer to the metal'' than even the standard
relational algebra, thanks again to the foundation of \noun{TAR}.
In our running example, variables like \lstinline!country! and \lstinline!keyword!
are backed directly by vectors in a column store, and relation combinators
compile to efficient operations over the columns. 

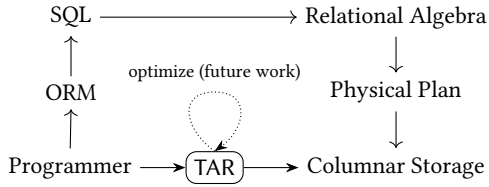
\begin{figure}
\begin{tikzpicture}
  \node[draw=none] (node2) at (-4.3,1.49) {SQL};
  \node[draw=none] (node1) at (-4.3,0.49) {ORM};
  \node[draw=none] (node6) at (-4.31,-0.5) {Programmer};
  \node[draw=none] (node3) at (0,1.49) {Relational Algebra};
  \node[draw=none] (node4) at (0,0.5) {Physical Plan};
  \node[draw=none] (node5) at (0,-0.5) {Columnar Storage};
  \draw[->] (node1.north) -- (node2.south);
  \draw[->] (node2.east) -- (node3.west);
  \draw[->] (node3.south) -- (node4.north);
  \draw[->] (node4.south) -- (node5.north);
  \draw[->] (node6.north) -- (node1.south);
  \node[draw, rounded corners] (node7) at (-2.4,-0.5) {\textsf{TAR}};
  \node[draw=none, node font=\footnotesize] (node8) at (-2.4,0.75) {optimize (future work)};
  \draw[densely dotted, arrows=-Stealth] (node7.north) .. controls (-2.8,0) and (-2.9,0.5) .. (-2.4,0.5) .. controls (-1.9,0.5) and (-2,0) .. (node7.north);
  \draw[arrows=-Stealth] (node6.east) -- (node7.west);
  \draw[arrows=-Stealth] (node7.east) -- (node5.west);
\end{tikzpicture}

\caption{\label{fig:compare-abstractions}Conventional database abstractions
(top route) v.s. unified abstraction based on \noun{TAR} (bottom route)}
\end{figure}

In summary, \figref{compare-abstractions} compares the architecture
of a traditional database with one based on \noun{TAR}. Unlike the
layers of abstractions that pull the programmer away from the data,
\noun{TAR} serves as the \emph{universal representation} shared by
the programmer, the storage engine, the execution engine, and the
optimizer, bringing each component closer to each other and enabling
cross-cutting optimizations. 

\section{Related Work\label{sec:Related-Work}}

\subsubsection*{\textquotedblleft What goes around comes around\textquotedblright}

Despite countless attempts at replacement, the relational model has
stood the test of time\,\citep{stonebraker2005what}. This paper
does not try to supplant the relational model, but rather to refine
and evolve it to meet modern needs. \noun{TAR}---despite being a
century older---can be understood as an evolution of the relational
algebra towards column stores and modern application code; the Prela
language can also be considered a prototype demonstrating how SQL
can be extended and made more flexible. In particular, \noun{TAR}
and Prela stand upon the two pillars of physical and logical independence:
the relation-centric \noun{TAR} assumes an abstract data model allowing
for different physical implementations including dense vectors, hash
maps, or bit sets; the corresponding Prela query accepts type annotations
to specify any of these data structures. Queries can be declared as
views and used as inputs to other queries, should the logical schema
change; addition and deletion of orthogonal attributes and relationships
also will not break existing queries. We believe ideas of \noun{TAR}
and Prela can be incorporated into mainstream relational databases
to simplify their design and implementation. 

\subsubsection*{TAR\noun{ }in theory}

The connection between Tarski and Codd has been noted as early as
the first \noun{PODS} conference in 1982, when Imielinski and Lipski\,\citep{DBLP:journals/jcss/ImielinskiL84}
demonstrated how to emulate Codd's relational algebra with Tarski's
\emph{cylindric algebras}\,\citep{henkin1971cylindric}, an extension
of Boolean algebras related to \noun{TAR}. Gyssens, Saxton and Van
Gucht\,\citep{DBLP:books/mk/freytagMV91/GyssensSG91} proposed an
alternative approach to modeling relational algebra by enhancing \noun{TAR}
with certain \emph{pairing/tagging} operations that introduce tuple
values. Prela's product operation \lstinline!.and! fills a similar
role to pairing and tagging, but is more natural for writing queries.
Database theorists have also used \noun{TAR} to analyze graph and
tree query languages like Regular Path Queries, XPath, and SPARQL\,\citep{DBLP:conf/icdt/FletcherGLBGVW11,DBLP:journals/jacm/LibkinMV16,DBLP:journals/igpl/SurinxFGLBGVW15,DBLP:journals/jlap/HellingsWGG22}.
This is thanks to how \noun{TAR}'s main operator, the relational composition,
conveniently models path traversal. Nevertheless, none of these languages
were built upon \noun{TAR} \emph{a priori}. Van den Bussche\,\citep{DBLP:conf/csl/Bussche01}
gives an introduction and survey of how Tarski's ideas have been applied
to database theory. 

\subsubsection*{\noun{TAR }in practice}

There have been sporadic efforts to bring the theory of \noun{TAR}
into practice. In 1992, Paredaens et al.\,\citep{DBLP:journals/sigmod/ParedaensBAGGTGSS92}
presented \noun{GOOD}, a ``Graph-Oriented Object Database'', and
envisioned an implementation over a data model based on \noun{TAR}.
The \noun{GOOD} system itself, however, was built atop a standard
relational engine. A follow-on technical report by Sarathy and Van
Gucht\,\citep{sarathy1993implementation} describes an implementation
of the \noun{GOOD} model in the \noun{IUGQL} query language. More
recently, Libkin et al. developed \noun{TriAL}\,\citep{DBLP:journals/tods/LibkinRSV18},
a language for recursive queries over RDF data based on a \emph{triple
algebra} related to \noun{TAR}. Compared to these languages, Prela
focuses on traditional non-recursive queries over relational data.
Prela also uses \noun{TAR} directly both as the surface syntax and
the underlying storage abstraction, whereas the aforementioned systems
all come with a separate, declarative query syntax. This indirection
complicates system design, obstructs performance tuning, and obscures
the beauty of \noun{TAR. }

\subsubsection*{Dataframes}

Dataframe libraries like \noun{pandas}\,\citep{mckinney2010pandas}
and \noun{polars}\,\citep{polars}\noun{ }provide an interface to
tabular data based on relational algebra. As such, these libraries
also struggle to express multi-table queries succinctly, as each join
operation must specify the join condition explicitly. At the engine
level, \noun{pandas} executes queries as-is, albeit \emph{eagerly}
-- every dataframe operation immediately produces a result when it
is called. Other more optimized engines like \noun{polars} generate
and optimize query plans, much like traditional OLAP databases. Prela's
engine can be thought of as implementing a lazy, push-based execution
model. Operators do not materialize intermediate results unless necessary,
and the unoptimized queries already run fast. 

\subsubsection*{SQL extensions}

The secret behind the longevity of SQL is its ability to absorb ideas
proposed in new query languages. Recent work by Shute, Zheng, and
Kudtarkar\,\citep{DBLP:conf/cidr/ShuteZK26} proposed SQL extensions
to better support semantic data modeling and graph workloads. Their
proposal lead to several design points strikingly similar to those
taken by Prela. For example, they also support implicit joins without
specifying join conditions, as well as multi-hop join/access syntax
similar to the last line of \figref{JOB-query-16b}. With Prela, we
aim to demonstrate that such extensions to SQL can be grounded on
the firm theoretical foundation of \noun{TAR}, and that they are not
merely syntactic sugar, but can produce tangible performance benefits
when connected to the underlying execution engine. 

\subsubsection*{Functional data models}

Several functional data models have been proposed over the years\,\citep{DBLP:conf/sigmod/BunemanF79,Gray2005FunctionalApproach,DBLP:journals/cj/KulkarniA86,DBLP:conf/sigmod/Shipman79},
and a recent position paper by Dittrich\,\citep{DBLP:conf/edbt/Dittrich26}
envisions a data model and query language where ``everything is a
function''. He then proposes a \emph{function algebra} where the
operators are higher-order functions consuming and producing functions.
Prela can be considered a realization of that vision over \noun{TAR}:
a relation in \noun{TAR} is precisely a (multi-valued) function. Prela
also goes beyond the syntax and semantics of the query language itself
and explores the synergy between high-level abstraction and low-level
execution. 

\section{The Prela Language}

This section defines the data model, syntax, and semantics of the
Prela query language. \Tabref{Relational-operators-in} provides a
reference of \noun{TAR} operators, their meaning, and their spelling
in Prela. 

\begin{figure}
\subfloat{%
\begin{tabular}{|c|c|c|c|}
\hline 
ID & title & year & keyword\tabularnewline
\hline 
\hline 
646 & The Godfather & 1972 & Crime\tabularnewline
\hline 
478 & Seven Samurai & 1954 & War\tabularnewline
\hline 
583 & Casablanca & 1942 & Romance\tabularnewline
\hline 
\end{tabular}}

\subfloat{%
\begin{tabular}{|c|c|}
\hline 
ID & \#\tabularnewline
\hline 
\hline 
646 & 0\tabularnewline
\hline 
478 & 1\tabularnewline
\hline 
583 & 2\tabularnewline
\hline 
\end{tabular} %
\begin{tabular}{|c|c|}
\hline 
\# & title\tabularnewline
\hline 
\hline 
0 & The Godfather\tabularnewline
\hline 
1 & Seven Samurai\tabularnewline
\hline 
2 & Casablanca\tabularnewline
\hline 
\end{tabular} %
\begin{tabular}{|c|c|}
\hline 
\# & year\tabularnewline
\hline 
\hline 
0 & 1972\tabularnewline
\hline 
1 & 1954\tabularnewline
\hline 
2 & 1942\tabularnewline
\hline 
\end{tabular} %
\begin{tabular}{|c|c|}
\hline 
\# & keyword\tabularnewline
\hline 
\hline 
0 & Crime\tabularnewline
\hline 
1 & War\tabularnewline
\hline 
2 & Romance\tabularnewline
\hline 
\end{tabular}}\caption{\label{fig:Decomposition-of-a}A wide table (top) and its decomposition
(bottom)}
\end{figure}

\subsection{Data Model\label{subsec:Data-Model}}

The fundamental datatype of Prela is the \emph{binary} relation. Although
restricting relations to be binary may appear limiting, it is simple
to recover the flexibility of multi-column tables with ones over only
two columns: every relation with $k$ columns is represented with
$k$ binary relations, each mapping the row number to the corresponding
column entry. \Figref{Decomposition-of-a} shows an example: a \lstinline!movie!
relation with columns \lstinline!title!, \lstinline!year!, and \lstinline!keyword!
decomposes to 4 binary relations: the first maps each ID to its row,
and the other 3 map each row to its title, year, and keyword, respectively.
The special treatment of the ID column will make sense when we join
across tables. 

Decomposing a wide table into many binary tables may appear expensive:
while the original table had $k$ columns, the decomposed tables now
span a total of $2k$ columns. However, since one column of each binary
relation simply contains the row number, that column need not be explicitly
stored, and each binary relation can be represented by a vector! In
\secref{Implementation} we will also show how Prela's implementation
eliminates the additional joins needed to ``re-assemble'' a wide
table from its columns. In other words, the ``binarization'' takes
a detour to arrive at the same physical representation found in column
stores. This detour is nevertheless necessary, as it allows us to
query the columns in a compositional way. 

\begin{table}
\caption{Relational operators in Prela\label{tab:Relational-operators-in}}
\begin{tabular}{l|l|l}
Name & Syntax & Semantics\tabularnewline
\hline 
\noun{compose} & \lstinline!r.select(t)! / \lstinline!r.s(t)! & $\pi_{r.1,t.2}(r\bowtie_{r.2=t.1}t)$\tabularnewline
\noun{product} & \lstinline!r.and(t)! & $r\bowtie_{r.1=t.1}t$\tabularnewline
\noun{predicate} & \lstinline!r.eq(v)!, \lstinline!r.lt(v)!, ... & $\sigma_{r.2=v}(r)$, $\sigma_{r.2<v}(r)$, ...\tabularnewline
\noun{filter} & \lstinline!r.filt(p)! & $\sigma_{p(r.2)}(r)$\tabularnewline
\noun{restrict} & \lstinline!r.with(t)!  & $r\ltimes_{r.2=t.1}t$\tabularnewline
\noun{map} & \lstinline!r.map(f)! & $\pi_{r.1,f(r.2)}(r)$\tabularnewline
\noun{gather} & \lstinline!r.gather(t)! & \emph{nested compose}\tablefootnote{Nested compose is not expressible in standard relational algebra;
formally, the semantics of \lstinline[basicstyle={\footnotesize\ttfamily}]!r.gather(t)!
is $\left\{ (x,\ddot{\bigcup}_{y\in r[x]}t[y])\mid x\in r.1\right\} $
where $\ddot{\bigcup}$ is the \emph{bag union} and $r[x]$ is the
bag of $y$ such that $(x,y)\in r$ (similar for $t[y]$). }\tabularnewline
\noun{invert} & \lstinline!r.inv()! & $\pi_{r.2,r.1}(r)$\tabularnewline
\noun{group by} & \lstinline!r.group_by(t)! & $\pi_{t.2,t.1}(t\ltimes_{t.1=r.2}r)$\tabularnewline
\end{tabular}
\end{table}

\subsection{Select, Project, and Join}

Prela queries are composed of operators applied to relational arguments.
The most important operator in Prela is the \emph{relational composition}
\lstinline!r.select(t)! which is equivalent to the relational algebra
expression $\pi_{r.1,t.2}(r\bowtie_{r.2=t.1}t)$, where $r.i$ indicates
the $i$-th column of $r$. Intuitively, relational composition generalizes
function composition in the same way (binary) relations generalize
functions: the composition $g\circ f$ is sometimes written $f;g$
and first applies $f$, then $g$. If the relation \lstinline!r!
maps each \lstinline!x! value to some \lstinline!y! values, and
if \lstinline!t! maps each \lstinline!y! to some \lstinline!z!s,
then \lstinline!r.select(t)! first maps \lstinline!x! via \lstinline!r!
to get \lstinline!y!s, then maps each \lstinline!y! via \lstinline!t!
to get \lstinline!z!s. Using our example in \figref{Decomposition-of-a},
if we call the binary relations \lstinline!movie!, \lstinline!title!,
\lstinline!year!, and \lstinline!keyword! from left to right, then
\lstinline!movie.select(title)! returns a binary relation mapping
each movie ID to its title, and similarly for \lstinline!movie.select(year)!
and \lstinline!movie.select(keyword)!. 

\begin{figure}
\begin{tabular}{|c|c|}
\hline 
\# & company\tabularnewline
\hline 
\hline 
0 & 657\tabularnewline
\hline 
1 & 188\tabularnewline
\hline 
2 & 353\tabularnewline
\hline 
\end{tabular}\hspace*{\fill}%
\begin{tabular}{|c|c|}
\hline 
ID & \#\tabularnewline
\hline 
\hline 
657 & 0\tabularnewline
\hline 
188 & 1\tabularnewline
\hline 
353 & 2\tabularnewline
\hline 
\end{tabular} %
\begin{tabular}{|c|c|}
\hline 
\# & name\tabularnewline
\hline 
\hline 
0 & Paramount\tabularnewline
\hline 
1 & Toho\tabularnewline
\hline 
2 & Warner Bros.\tabularnewline
\hline 
\end{tabular} %
\begin{tabular}{|c|c|}
\hline 
\# & country\tabularnewline
\hline 
\hline 
0 & USA\tabularnewline
\hline 
1 & Japan\tabularnewline
\hline 
2 & USA\tabularnewline
\hline 
\end{tabular}

\caption{Left: decomposed \lstinline!company! column from the \lstinline!movie!
table; right: decomposed columns from the \lstinline!movie_company!
table\label{fig:Left:-decomposed-}}
\end{figure}

Composition also plays the role of joins in Prela. Consider a \lstinline!movie_company!
table with three columns: the company ID, name, and country, each
of which becomes a binary relation in Prela as shown on the right
of \figref{Left:-decomposed-}. We also add a \lstinline!company!
column to the \lstinline!movie! table linking each movie to the ID
of its production company. Now, we can look up the country of a movie's
production company with \lstinline!movie.s(company).s(id2row).s(country)!,
where \lstinline!.s! is shorthand for \lstinline!.select!, and \lstinline!id2row!
is the relation mapping company IDs to row numbers. Because joining
via foreign keys almost always require ``resolving'' the key to
its row, Prela automatically inserts \lstinline!.s(id2row)! to key/foreign
key joins, similar to how programming languages like Rust automatically
dereference pointers upon a field access. This allows us to simply
write \lstinline!movie.s(company).s(country)! which is pronounced
``movie's company's country''. This is also what happened on the
last line of \figref{JOB-query-16b}. 

Given that composition ``throws away'' the join attributes, the
reader may be wondering if there are certain SQL queries that Prela
cannot express. Our answer is two-fold. On one hand, real world queries
almost always join on key/foreign keys, and keys rarely serve any
other purpose --- it does not make sense to aggregate over them.
So dropping the keys after the join is not a real sacrifice. On the
other hand, it is true that the original \noun{TAR} is strictly less
expressive than Codd's relational algebra. One approach to bridging
this gap is to introduce the so-called \emph{pairing operators} to
\noun{TAR}\,\citep{DBLP:books/mk/freytagMV91/GyssensSG91}. In Prela,
we propose the\emph{ product} operator \lstinline!.and! as an equivalent
yet more practical extension. \lstinline!r.and(t)! joins the relations
on their first column, i.e., $r\bowtie_{r.1=t.1}t$. Crucially, the
output is still a \emph{binary} relation, whose first column is the
intersection of $r.1$ and $t.1$, and the second column is the Cartesian
product of $r.2$ and $t.2$ for each matching $r.1=t.1$. For example,
\lstinline!title.and(year)! is the relation mapping 0 to (The Godfather,
1972), 1 to (Seven Samurai, 1954), and 2 to (Casablanca, 1942). Combining
with \lstinline!.select!, the expression \lstinline!movie.select(title.and(year).and(keyword))!
implements the SQL query \lstinline[language=SQL]!SELECT title, year, keyword FROM movie!. 

The next Prela operator exercised by \figref{JOB-query-16b} is the
\emph{predicate} \lstinline!r.eq(v)! which returns all rows of \lstinline!r!
whose right column equals \lstinline!v!. Other predicates like \lstinline!.gt!
(greater than), and \lstinline!.rx! (regular expression match) work
the same way. There is also a generic \lstinline!r.filt(p)! operator
that filters \lstinline!r! by any Boolean function \lstinline!p!.
Going back to the example, \lstinline!keyword.eq("character-name-in-title")!
returns rows in \lstinline!keyword! (which maps a movie to its keywords)
such that the keyword is ``character-name-in-title''; \lstinline!company.s(country).eq("[us]")!
first composes \lstinline!company! (mapping a movie to its company)
with \lstinline!country! (mapping a company to its country) to get
a mapping from each movie to its country, then keeps only those in
the US. Finally, the \lstinline!.and! operator joins the filtered
relations, keeping only American movies with character name in their
titles. 

The last operator used in our example is the \emph{restriction} \lstinline!r.with(t)!
which is exactly the left-semijoin $r\ltimes_{r.2=t.1}t$, and it
emulates SQL's \lstinline[language=SQL]!WHERE! clause.\footnote{Unfortunately \lstinline!where! is a reserved keyword in Rust.}
A happy accident is that \lstinline!.and! doubles as ``logical conjunction''
inside \lstinline!.with!, because the semijoin ignores the second
column of its right-hand-side. 

\begin{figure}
\begin{minipage}[t]{0.55\columnwidth}%
\begin{lstlisting}[language=SQL,deletekeywords={year, YEAR},deletekeywords={[2]year, [2]YEAR}]
SELECT keyword, min(year)
  FROM movie
 GROUP BY keyword
HAVING min(year) > 1950
\end{lstlisting}
\end{minipage}%
\begin{minipage}[t]{0.45\columnwidth}%
\begin{lstlisting}
movie.group_by(keyword)
     .gather(year)
     .map(min)
     .gt(1950)
\end{lstlisting}
\end{minipage}

\caption{Grouping and aggregation in SQL and Prela\label{fig:Grouping-and-aggregation}}
\end{figure}

\subsection{UDFs, Structured Output, and Aggregation\label{subsec:Grouping-and-Aggregation}}

Just like the original form of Codd's relational algebra, \noun{TAR}
was designed to model first order logic, and requires extensions to
support practical workloads. Prela supports ``generalized projection''
with the \lstinline!.map! operator: \lstinline!r.map(f)! maps the
function \lstinline!f! over each value in the second column of \lstinline!r!.
One advantage of embedding Prela in Rust is to allow arbitrary Rust
code to be used as UDFs. In particular, the query can construct structured
outputs. For example, the following query returns a list of \lstinline!Movie!
structs instead of tuples:

\begin{lstlisting}
movie.select(title.and(year))
     .map(|(id,(t,y))| Movie{title: t, year: y})
\end{lstlisting}
Another extension allows Prela to return \emph{nested} outputs: \lstinline!r.gather(t)!
is like \lstinline!r.select(t)!, but maps each distinct \lstinline!r.1!
to a list of \lstinline!t.2! values. For example, \lstinline!movie.gather(keyword)!
returns a binary relation mapping each movie ID to all the keywords
associated with that movie. Together with \lstinline!.map!, we can
use \lstinline!.gather! to implement aggregation: \lstinline!movie.gather(year).map(min)!
returns the minimum year associated with each movie, by mapping the
list function \lstinline!min! over the list of years per movie. 

To support grouping, we need to introduce one last operation from
the original \noun{TAR}: the \emph{inverse} \lstinline!t.inv()! simply
flips the two columns of \lstinline!t!. We can now implement Prela's
\lstinline!r.group_by(t)! as \lstinline!t.inv().with(r.inv())!,
or $\pi_{t.2,t.1}(t\ltimes_{t.1=r.2}r)$ in relational algebra. For
example, the first line of the Prela query in \figref{Grouping-and-aggregation}
desugars to \lstinline!keyword.inv().with(movie.inv())!. First, \lstinline!keyword.inv()!
flips the \lstinline!keyword! relation to map each keyword to the
rows it appears in; restricting with \lstinline!movie.inv()! is a
no-op, but would be meaningful if \lstinline!movie! were already
restricted by other predicates. Line 2 composes the result from line
1 with \lstinline!year! to obtain a mapping from keyword to years.
Line 3 aggregates with \lstinline!min! to find the earliest year
associated with each keyword. Finally, the last line \lstinline!.gt(1950)!
implements the \lstinline[language=SQL]!HAVING! clause in the SQL
query. Grouping by multiple columns can be achieved with the help
of \lstinline!.and!, for example \lstinline!movie.group_by(keyword.and(year))!.
To aggregate over multiple columns, we can pass in a function that
reduces all of them at once. 

\subsection{Discussion}

Looking back at the example in \figref{JOB-query-16b}, we see another,
more intuitive way to understand the query: pretend that there are
\lstinline!movie! objects with field \lstinline!company!, \lstinline!keyword!,
\lstinline!title!, and \lstinline!cast!, where \lstinline!company!
is its own object with field \lstinline!country!, and similarly for
\lstinline!cast!, \lstinline!person!, and \lstinline!alias!. Then
the Prela query appears to access fields and apply predicates to them.
This is not accidental. Central to the design of both \noun{TAR} and
Prela is the focus on binary relations which closely model both attributes
and relationships \emph{à la} Entity/Relationship. Modern business
logic usually start life as an E/R diagram, and the application code---commonly
written in some object-oriented language---naturally mirrors the
E/R model. Built with a set of compositional operators, Prela queries
blend into the surrounding application, obviating the need for ORMs. 

\section{Implementation\label{sec:Implementation}}

In \Subsecref{Data-Model} we explained how the apparent storage overhead
of binary relations is eliminated by columnar storage. It may appear
a computational overhead remains, when we need to \emph{reassemble}
a table from its columns. Consider the following SQL query:

\begin{lstlisting}[language=SQL]
SELECT ID, title, keyword FROM movie
\end{lstlisting}
It outputs the ID, title, and keyword of each movie. In a conventional
database the query requires only a simple scan over the \lstinline!movie!
table. Specifically, the execution over a column store would look
something like this: 

\begin{lstlisting}[language=Python]
for i in 0..n: print(movie[i],title[i],keyword[i])
\end{lstlisting}
In the above, a single loop \emph{co-iterates} over the selected table
columns. The corresponding Prela query is \lstinline!movie.select(title.and(year))!.
Implemented naively, the Prela query would require two joins, one
each for \lstinline!.and! and \lstinline!.select!. In general, the
number of joins grows proportionally with the number of columns involved
in a Prela query, which is rather expensive. 

The textbook solution to remove intermediate state during query execution
is the iterator model\,\citep{DBLP:journals/tkde/Graefe94}. Iterators
incur overhead like method calls and other bookkeeping operations,
and two standard remedies are vectorization and query compilation\,\citep{DBLP:journals/pvldb/KerstenLKNPB18}.
Unfortunately, both approaches require significant engineering effort
that would exceed the scope of a prototype like Prela. To achieve
good performance while keeping Prela's implementation simple and modular,
we turn to a technique from functional compilers --- deforestation
in continuation-passing style (CPS)\,\citep{DBLP:conf/fpca/GillLJ93}. 

\begin{figure}
\begin{lstlisting}
lhs.select(rhs).drive(k) =
    lhs.drive(|x, y| rhs.probe(y, |z| k(x, z)))
lhs.and(rhs).probe(x, k) =
    lhs.probe(x, |y| rhs.probe(x, |z| k((y, z))))
col.drive(k) = for i in 0..col.len: k(i, col[i])
col.probe(i, k) = k(col[i])
\end{lstlisting}
\caption{\label{fig:CPS-protocol}CPS protocol for \lstinline!.select!, \lstinline!.and!,
and columns}
\end{figure}

The basic idea of CPS is for combinators to push computation down
from the query root to its leaves instead of pulling data up from
leaves to root. More specifically, Prela queries can be accessed in
two ways. Calling \lstinline!query.drive(k)! applies the closure
\lstinline!k! (the \emph{continuation}) to each row of \lstinline!query!.
Calling \lstinline!query.probe(key, k)! looks up \lstinline!key!
in \lstinline!query! and applies \lstinline!k! to each resulting
value. For example, \lstinline!movie.s(title).drive(print)! prints
each \lstinline!(id, title)! pair, while \lstinline!movie.s(title).probe(3, print)!
prints the title of the movie with ID 3. Since every combinator receives
and returns binary relations, \lstinline!.drive! and \lstinline!.probe!
are well-defined for all Prela queries. \Figref{CPS-protocol} shows
the definition of \lstinline!.drive! and \lstinline!.probe! for
\lstinline!.select!, \lstinline!.and!, as well as input columns. 

Let us unfold the definitions for our example query. We start by driving
the query with a continuation \lstinline!k! which may print each
result, or append it to a buffer:

\begin{lstlisting}[language=Prela]
movie.select(title.and(keyword)).drive(k)
\end{lstlisting}
To save space we will abbreviate \lstinline!movie!, \lstinline!title!,
\lstinline!keyword!, \lstinline!.drive!, and \lstinline!.probe!
with \lstinline!m!, \lstinline!t!, \lstinline!w!, \lstinline!.drv!,
and \lstinline!.prb!, respectively. Expanding \lstinline!.drive!
on \lstinline!.select! drives its left-hand-side and probes its right-hand-side:

\begin{lstlisting}[language=Prela]
m.drv(|x, y| t.and(w).prb(y, |z| k(x, z)))
\end{lstlisting}
As \lstinline!m! is an input column, driving it expands to a loop: 

\begin{lstlisting}
for i in 0..m.len: t.and(w).prb(i, |z| k(m[i], z))
\end{lstlisting}
Next, probing \lstinline!.and! propagates to probing its arguments:

\begin{lstlisting}
for i in 0..m.len:
    t.prb(i, |y| w.prb(i, |z| k(m[i], (y, z))))
\end{lstlisting}
Finally, probing \lstinline!title! and \lstinline!keyword! expand
to array accesses:

\begin{lstlisting}
for i in 0..m.len: k(m[i], (t[i], w[i]))
\end{lstlisting}
At this point, we recover exactly the fused co-iteration shown earlier
in this section! Note how we use the term \emph{expand}: in Prela's
implementation, all we had to do is annotate the definition of \lstinline!.drive!
and \lstinline!.probe! for each combinator with \lstinline!#[inline]!,
and the Rust compiler would automatically fuse the operations into
tight loops. 

The main motivation for implementing Prela in continuation-passing
style is to derive a self-contained, easy-to-understand prototype
that can guide the construction of more full-fledged systems. On one
hand, relying on the Rust compiler has unacceptably high latency for
queries that must be compiled on-the-fly; on the other hand, data
warehouses repeatedly execute the same query over gradually-changing
data, which can amortize the compilation cost\,\citep{DBLP:journals/pvldb/RenenHPVDNLSKK24}.
Another advantage of embedding Prela in Rust is to allow arbitrary
Rust code to be used as UDFs, as described in \subsecref{Grouping-and-Aggregation}.
We pose a challenge to future research to match Prela's flexibility
and run time performance, while incurring much lower compilation overhead. 

\section{Evaluation}

In \secref{Introduction} we promised that Prela leads to concise,
clear, and efficient queries. Clarity is subjective, and measuring
conciseness in lines of code only tells part of the story. We therefore
invite the reader to compare Prela queries to their SQL counterparts
to judge for themselves.\footnote{The source code for Prela and example queries are available at \url{https://prela-lang.org}.}
Nevertheless, we share some ``quick numbers'' here: the Prela code
for all 113 queries in the Join Order Benchmark\,\citep{DBLP:journals/pvldb/LeisGMBK015}
(JOB) total \textasciitilde 1000 lines, while the original SQL queries
span over 4000 lines; for TPC-H\,\citep{tpch}, it takes Prela \textasciitilde 300
lines to express all 22 queries, whereas the SQL queries span \textasciitilde 600
lines. Prela achieves more savings on JOB because the queries there
join together many tables, and the implicit/structural joins in Prela
avoids spelling out the join conditions, as demonstrated in \figref{JOB-query-16b}. 

Does the brevity of Prela queries sacrifice their performance? To
answer this question, we benchmark Prela against DuckDB v1.5.3 on
JOB and TPC-H. Care must be taken to interpret our results. On one
hand, Prela is an early prototype and has not implemented the myriad
of optimizations found in mature systems; on the other, Prela is also
not weighed down by a large set of features. In addition, Prela has
no query optimizer, but offers more flexibility for performance tuning.
One way to understand our experiments is that they gauge the \emph{performance
ceiling} of a database system built on a foundation of \noun{TAR},
assuming such a system implements a competent query optimizer as well
as low-level optimizations. 

All experiments run on an M2 MacBook Air with 16 GiB of RAM. Data
is pre-loaded into memory. Since Prela is currently single-threaded,
we also run DuckDB on a single thread for fairness. Each query is
executed 5 times and the median run time is reported. 

\begin{figure}
\includegraphics[width=0.5\columnwidth]{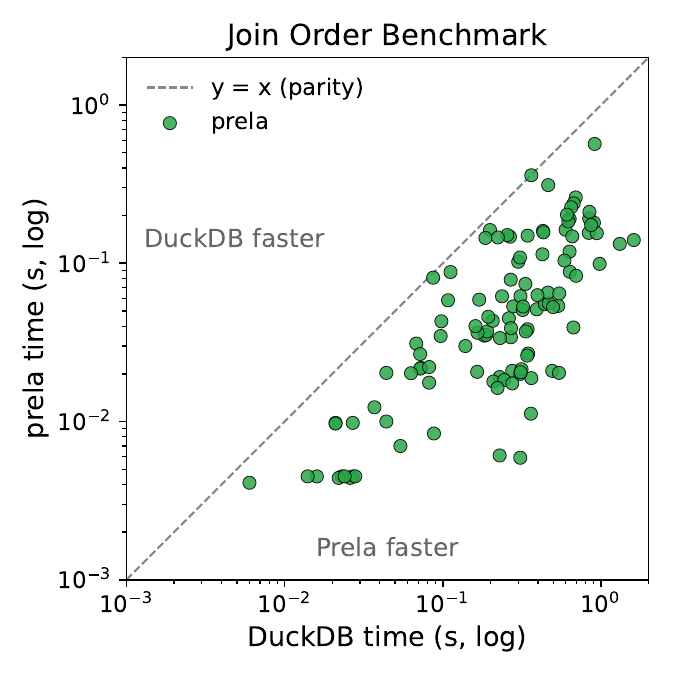}\includegraphics[width=0.5\columnwidth]{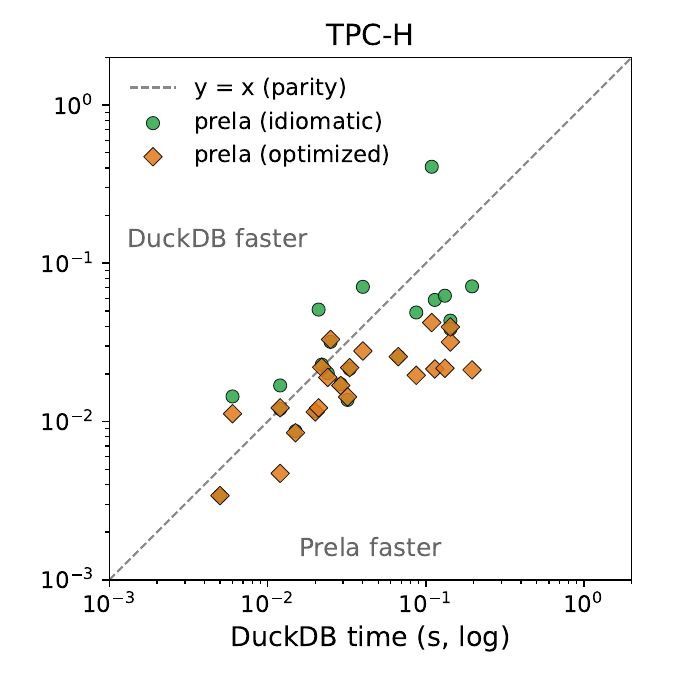}

\caption{\label{fig:perf_plot}Performance comparison of Prela and DuckDB:
each data point represents a query, where the $x$-coordinate is DuckDB's
run time, and the $y$-coordinate is Prela's run time}
\end{figure}

Summaries of the results are collected in \figref{perf_plot}. Prela
executes all 113 queries of JOB in 8.68s, a 4.5$\times$ speedup over
DuckDB's 38.82s. Prela's advantage is mainly due to data representation:
because JOB's primary keys are contiguous integers, the ID to row
mapping degenerates to the identity after sorting; instead of materializing
the relation, we represent this mapping with the identity function
which is optimized away during compilation. And because each column
is stored as a vector indexed by the row number, joins degenerate
to fast array accesses. We tried to port the same optimizations back
to DuckDB but were not successful. The only user-accessible index
provided by DuckDB is the hashing-based adaptive radix tree\,\citep{DBLP:conf/icde/LeisK013},
and its query optimizer ignored every foreign key index. DuckDB frequently
picks bushy join plans, whereas Prela follows left-deep plans shaped
by the query structure and uses index nested loop joins to avoid allocation.
For TPC-H, the plot shows two sets of queries: the circular dots represent
\emph{idiomatic} queries which are written in the most natural way
without regard to performance. These queries run in 1.13s, which is
1.11$\times$ faster than DuckDB's 1.26s. It is easy to further improve
the performance of the Prela queries. By directly rewriting the queries,
we were able to speed them up by reordering joins, introducing materialization
points, and specifying data structures like bit sets for materialized
intermediates. The tuned queries take 0.44s, 2.86$\times$ faster
than DuckDB. 

The experiments show how Prela empowers the programmer to fine tune
query performance, and a future query optimizer can achieve results
on par or exceeding that of current systems. 

\section{Conclusion}

We have presented the design and implementation of Prela, a query
language based on \noun{TAR}. At the core of Prela is the binary relation
which connects columnar storage with entity/relationship semantics,
resulting in succinct and efficient queries. Our experiments show
there is strong potential for a database system based on \noun{TAR}
to outperform state-of-the-art OLAP systems. Future work include parallelization,
fast compilation, and query optimization of \noun{TAR}, as well as
consideration of transactional workloads. 

\balance

\bibliographystyle{ACM-Reference-Format}
\bibliography{references}

\end{document}